**Title**:
An efficient dosimetry method with a Faraday cup for small animal, small-field proton irradiation under conventional and ultra-high dose rates

**Authors**:
Abbas Husain*[1], Jeremiah Ryser*[1], Julia Pakela[1], Menggui Huang[1], Khayrullo Shoniyozov[1], Francois V. Stappen[2], Costas Koumenis[1], Lei Dong[1], Yi Fan[1], Eric Diffenderfer[1], Wei Zou[1]

1. Department of Radiation Oncology
   University of Pennsylvania
   Philadelphia, PA
   wei.zou@pennmedicine.upenn.edu
2. IBA-PT
   Philadelphia, PA

**Running Title:**
Small-field Proton FLASH dosimetry

**Author Contribution Statement:**
W. Zou, E. Diffenderfer, M. Huang, Y. Fan worked on the study conception and design; E. Diffenderfer and F. Stappen worked on UHDR beam generation and control; A. Husain, J. Ryser, K. Shoniyozov and W. Zou performed the data collection and analysis; A. Husain, J. Ryser, J. Pakela, W. Zou drafted the manuscript; all authors reviewed and approved the results and final manuscript.

**Conflict of Interest:**
This work is partially supported by NIH P01CA257904-01A1 grant. Additionally, Francois Vander Stappen is employed by iba Proton Therapy.

**ABSTRACT:**

**Introduction**:

We developed and evaluated a method for dose calibration and monitoring under conventional and ultra-high dose rates for small animal experiments with small-field proton beams using a Faraday cup.

**Methods**: We determined a relationship between dose and optical density (OD) of EBT-XD Gafchromic™ film using scanned 10x10 cm$^2$ proton pencil beams delivered at clinical dose rates; the dose was measured with an Advanced Markus chamber. On a small animal proton irradiation platform, double-scattered pencil beams with 5 or 8 mm diameter brass collimation at conventional and ultra-high dose rates were delivered to the EBT-XD films. The proton fluence charges were collected by a Faraday cup placed downstream from the film. The average of the irradiated film ODs was related to the Faraday cup charges. A conversion from the Faraday cup charge to the average dose of the small-field proton beam was then obtained.

**Results**: The relationship between the small-field average profile dose and Faraday cup charge was established for 10 and 15 Gy mice FLASH experiments. The film OD was found to be independent of dose rate. At small-animal treatments, the Faraday cup readings were conveniently used to QA and monitor the delivered dose and dose rates to the mice under conventional and ultra-high dose rates.

**Conclusion**: The dose calibration and monitoring method with Faraday cup for small animal proton FLASH experiments is time-efficient and cost-effective and can be used for irradiations of various small field sizes. The same approach can also be adopted for clinical proton dosimetry for small-field irradiations.

**INTRODUCTION**

Radiation therapy delivered at ultra-high dose rates has shown potential for enhanced sparing of normal tissues while maintaining local tumor control, an effect named FLASH.[1–3] While the biological mechanism(s) behind the FLASH effect is not fully understood, it is theorized that oxygen depletion under conditions of ultra-high dose rates plays an important role in normal tissue sparing.[4, 5] It is hypothesized that FLASH irradiation causes depletion of oxygen faster than it can be replenished, leading to acute hypoxia and transient radio-resistance in normal tissues.[6] This could allow for sparing of naturally hypoxic cell types, such as those found in stem cell niches in normal tissues.[7] Another potential pathway for the FLASH effect is immune sparing. Because FLASH irradiation occurs on a timescale of milliseconds, the number of immune cells circulating through the path of radiation is minimal in comparison with conventional fractionation. It is hypothesized that this immune sparing may result in a stronger immune boost post-radiation.[8] Further pre-clinical studies are necessary to fully understand the mechanisms behind the FLASH effect and also to identify the treatment conditions (including dose, dose rate, fractionation schedule, and treatment site) that could provide clinical benefits for human patients. This necessitates an efficient dose monitoring system that can function under high dose rates for small-fields.

Previous small animal studies using proton FLASH irradiation have utilized scatterers to spread out the beam laterally and achieve a mostly uniform fluence prior to irradiating the target.[8] However, a completely uniform field is difficult to achieve, especially under the further collimation requirded with small animal targets. The proton FLASH mouse experiments in this study require 5 and 8 mm collimations for mice medulloblastoma brain tumor irradiations. As Gottschalk[9] has described, for a single pencil beam, an ion chamber with a large cross-section encompassing the entire beam is needed to monitor the integral dose. The same applies to scattered and collimated proton beams. However, under ultra-high dose rates, the collectoin efficiency of ion chambers suffers due to ion-recombination effects, which need to be characterized and corrected, if necessary, under applied bias voltage and beam currents[10, 11]. Diode detectors typically used for small-field dosimetry have small active areas (~5 mm) which are sensitive to small setup shifts, making them impracticable when dealing with a non-uniform small field. Radiochromic film avoids this issue by capturing the entire field and has been found to be useful for dosimetry measurements in small-field proton beams due to its high resolution in 2D.[12, 13] Applications of film dosimetry also extend to other high dose rate modalities[14, 15] where the optical density (OD) is found to be dose-rate independent[16, 17]. Radiochromic film, however, suffers from high cost and time-consuming readout process. Additionally, readouts from radiochromic film can display sensitivity to a number of variables including film orientation, history, and batch number, leading to potential inconsistencies across measurements.[18, 19]

Faraday cups (FC) have been used in proton measurements for over 40 years due to their dose-rate independence.[20, 21] Faraday cup measurements are made in air and are based on proton fluence, giving an absolute measurement of fluence for proton beams.[22] Goma et al.[23] compared two approaches, absolute dosimetry with a Faraday cup and reference dosimetry with an ionization chamber, under clinical dose rate PBS beams and concluded equivalency of both approaches when a plane-parallel ionization chamber is used in the reference dosimetry. In scanned pencil beams, Faraday cups were found not to suffer from recombination effects or saturate and approach unity collection efficiency[24, 25]. Its independence of dose rate[22, 25, 26], even up to high dose rates[10, 27], provides reproducible measurements with good accuracy for proton irradiations. This study presents the design and evaluation of a dosimetry method for small-animal studies with a double scattered small-field proton beam using radiochromic film as an intermediary to relate charge collected in a Faraday cup to absolute dose. Conventional and ultra-high dose-rates were considered.

**MATERIALS AND METHODS**

**Dose to film OD calibration**

To determine the relationship of the dose to film optical density (OD), radiochromic film measurements were taken under uniform proton fields from clinical scanning beams and compared against the dose readouts from an ADCL-calibrated parallel-plate Advanced Markus ion chamber (PTW, Germany). EBT-XD Gafchromic™ film (Ashland, NJ) was placed on top of the parallel-plate ion chamber positioned at the isocenter of the beam (Figure 1). A 2 cm solid-water buildup was placed on top of the film to shift the measurement beyond the depth-dose buildup region and into the plateau region.[28, 29] This film-type was selected due to its ability to self-develop, as well as its sensitivity and advantageous film reading characteristics in high dose regions,[30, 31] as the mice FLASH experiments typically require about or above 10 Gy irradiations. Films of 2x2 cm$^2$ were cut out from one original large film sheet and placed at the same orientation intercepting the beams to avoid potential readout variations due to film lot and orientations. Scanned 10x10 cm$^2$ proton pencil beams of 225 MeV were delivered at various MU per spot at 4 mm spacing from a clinical treatment beamline (Proteus Plus, IBA, Belgium). The fields were uniform and the dose ranged from about 5 to 20 Gy. Each field was repeatedly delivered two times to separate films and the corresponding ion chamber dose readouts were collected.

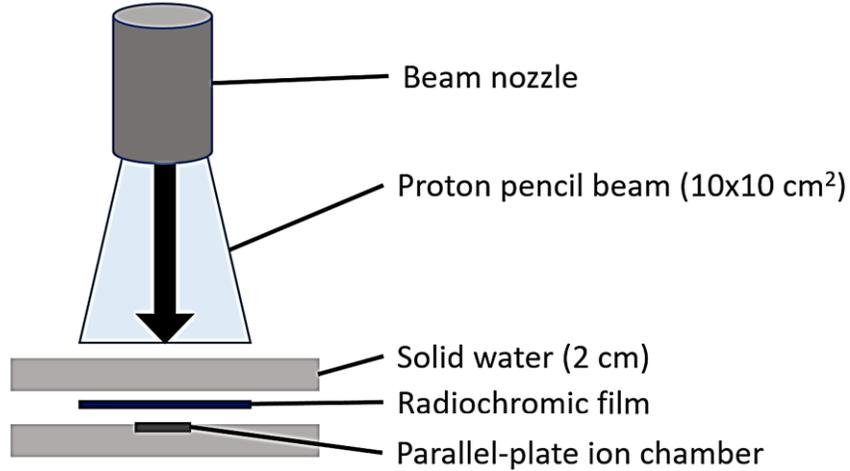

**Fig. 1.** Proton irradiation setup with clinical pencil beam scanning fields for obtaining film OD and dose relationship.

The response of the Gafchromic EBT-XD film to radiation was quantified through optical density (OD), defined as a function of light transmittance through film at an irradiated region of interest, I, and an unirradiated background region, $I_0$:

$$OD = -\log \frac{I}{I_0}.$$

Films were scanned at 300 dots-per-inch (DPI) in .tiff format. The OD in the region of interest was calculated from the red-channel transmittance values using an in-house MATLAB script. Mean OD within an arbitrarily identified 1x1 cm² irradiated region on the film was calculated. The relationship between the film OD values and the chamber dose readouts was fitted to a third-order polynomial,

$$OD = f(Dose) = A * Dose^3 + B * Dose^2 + C * Dose + D \qquad (1)$$

where *A, B, C,* and *D* are fitted polynomial coefficients.

The small-field, small-animal irradiation setup used in this study is shown in Figure 2. A horizontal pencil beam at 227 MeV at either low or ultra-high beam intensity was delivered through the beamline to the small animal research room.[8] Here it was scattered and then collimated to the desired field size. The first scatterer (1 mm Pb foil) spread out the beam laterally, and the second scatterer (5 mm Pb beads) improved fluence uniformity. The beam then passed through a brass collimator with an aperture of 5 or 8 mm in diameter before irradiating either the film (for calibration) or the mouse (for experiments). Downstream was a Faraday cup (BC-75, Pyramid Technical Consultants, MA) with a 7.5 cm effective collection diameter that collected the proton fluence. The FC charge was readout by an electrometer (Dose1, IBA, Belgium) with 0 V bias. During calibration and animal experiments, solid water was used for buildup to irradiate at the depth-dose plateau region. Films and FC readings were collected from irradiations with various durations at conventional and ultra-high dose rates. These irradiations generated the film ODs at a similar range to the ones from the scanned 10x10 cm$^2$ fields with the clinical beams.

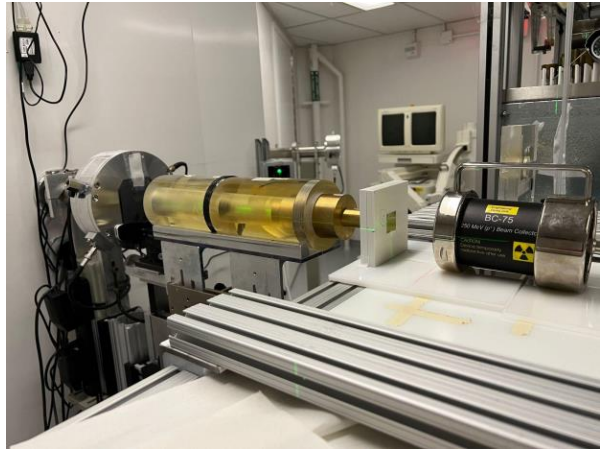

**Fig. 2.** The small-field, small-animal proton irradiation beam setup for calibration. The beam was double scattered and passed through a brass aperture for collimation. 2 cm of solid water buildup was used before film irradiation and the proton fluence was collected by the Faraday cup. The 227 MeV beam was generated by IBA Proteus Plus and delivered at both conventional and ultra-high dose rates.

The film ODs were obtained through similar approach as above. The film pixels in the circular irradiated region were identified and the mean OD in this region was obtained. The mean OD profile was related against the Faraday cup charge with a third-order polynomial fit to the data,

$$FCC = g(OD) = E*OD^3 + F*OD^2 + G*OD + H \qquad (2)$$

where *E, F, G,* and *H* are fitted polynomial coefficients. Finally, this function (Eqn. 2) was combined with the previous relationship between dose and film OD (Eqn. 1) to obtain a conversion from the Faraday cup charge to the average dose of the small-field proton beam as,

$$FCC = g(OD) = g(f(Dose)). \qquad (3)$$

**RESULTS**

Proton pencil-beam-scanning fields of 10x10 cm$^2$ with 8, 10, 12, 15, 18, 20, 25 and 30 spot MUs at 4 mm spacing were delivered to films and Advanced Markus chamber in clinical treatment rooms. The physical dose readouts from the ion chamber varied from 5.35 to 20.08 Gy. The mean film OD values were plotted against the dose measured from the ion chamber (Figure 3). A 3$^{rd}$-order polynomial fitting the OD vs. dose (Eqn. 1) was obtained with $R^2$ of 0.9998.

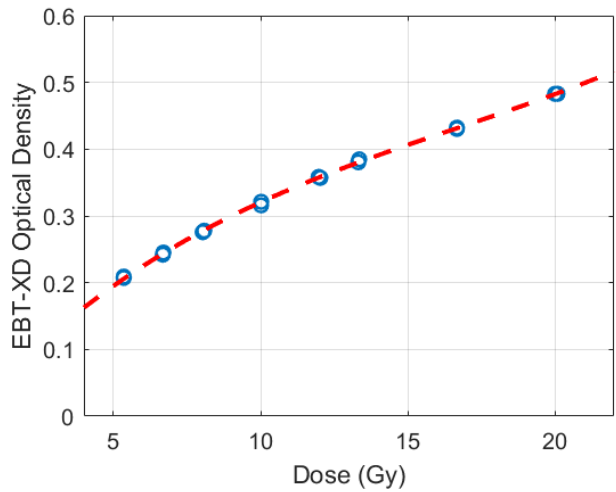

**Fig. 3.** Dose vs. Gafchromic™ EBT-XD film optical density plot for scanned proton fields. The dashed line represents the polynomial fit to Eqn. 1.

Using the Gafchromic EBT-XD films cut from the same sheet, the film OD vs. Faraday cup readings for a diameter of 5 mm collimated beams at conventional and ultra-high dose rates were obtained (Figure 4). The data from beam deliveries at both dose rates fell onto the same curve, which indicates that EBT-XD film has the same dose response independent of the proton dose rates. A 3$^{rd}$-order polynomial fitting the FC reading and the OD (Eqn. 2) was obtained for the delivery at both dose rates, with $R^2$ of 0.9921. The same procedure was repeated for a diameter of 8 mm collimated beams.

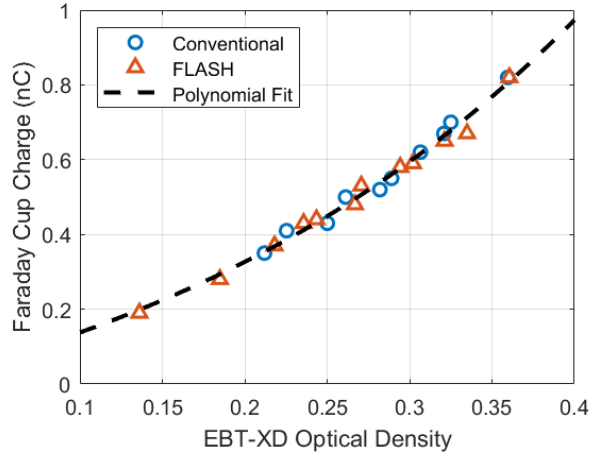

**Fig. 4.** Film net optical density vs. Faraday cup charge for conventional and ultra-high dose rates for 5 mm collimated beams. The dashed line represents a polynomial fit to Eqn. 2.

Combining Eqn. 1 and 2 into Eqn. 3, the calibration curve of the mean irradiated dose versus the Faraday cup charge can be obtained. Although Eqn. 3 is complicated in its form, the FC charge appears linear with the delivered dose (Figure 5), which is expected. Based on this calibration curve, for mice experiments to deliver 10 and 15 Gy, the FC charge shall be 0.645 and 0.964 nC. To verify the results, we tuned the beams to deliver the FC charge to the desired charges and two films for each setting were collected. The mean ODs of the irradiated region indicates dose deliveries of 10.07, 10.07, 15.09 and 14.93 Gy.

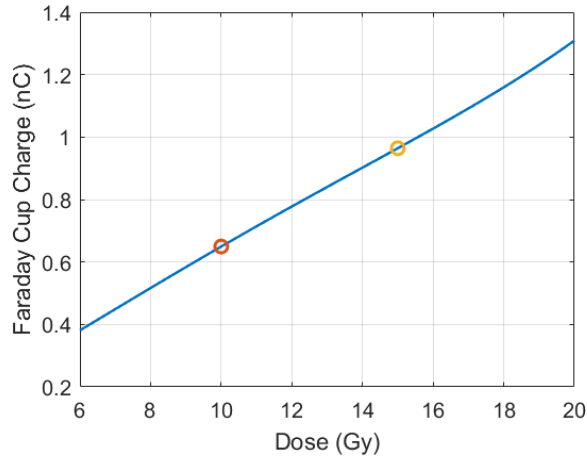

**Fig. 5.** Plot of Faraday cup charge vs. dose from Eqn. 3. Marked circles are the FC readings for dose 10 and 15 Gy deliveries.

In our proton small animal irradiation system, an oscilloscope is connected with the beam control unit and used to measure the delivery duration of cyclotron accelerated proton beams[8]. Such time information combined with FC charge readings can yield the delivery dose rate, where the pencil beam is considered quasi-continuous[10]. At the beginning of each mouse experiment, quality assurance (QA) was performed with the FC to tune the duration of the beam to deliver the desired FC charge. Dose rate was also checked with the oscilloscope readings as part of the QA procedure.

Experiments were performed on mice bred with medulloblastoma brain tumors. The juvenile mouse brains were irradiated with the same beam duration in order to treat with the same dose. The proton beams shoot through the juvenile mice cerebellum regions and the fluence was collected by the FC. For 60 juvenile mice irradiated with desired 10 Gy to ultra-high dose rate (37 mice) and conventional dose rate (23 mice), the monitored dose with Faraday cup was found to be 10.02±0.13 Gy. The ultra-high dose rate was found to be 129.7±5.3 Gy/s; the conventional dose rate was found to be 0.75±0.04 Gy/s.

**DISCUSSION**

We have established a dose calibration and monitoring method for small animal proton FLASH experiments with small-field irradiation by relating OD in radiochromic EBT-XD film to dose measured in a parallel-plate ion chamber, and then Faraday cup charge to OD. The benefit of this approach is that once a calibration curve has been established, as shown in Figure 5, the Faraday cup readings can be used to relate to various required dose deliveries. Since the film and Faraday cup readings are both independent of the dose rates as shown in this study and other literatures[10, 27], the same relationship is applied to both conventional and ultra-high dose rate irradiations. The read out of the FC by an electrometer is instantaneous and reliable. With an oscilloscope to record the irradiation duration, the dose rate can be conveniently obtained. For different collimated field sizes, new calibration curves for FC charges vs. dose need to be established as the proton fluence varies and the beam cross-profile is not flat.

Film, on the other hand, can also be used for dose and dose rate (with delivery duration information) monitoring purposes, once a dose vs. OD calibration curve is established, as in Figure 3. However, this process requires re-entering the treatment room and placement of film for each irradiation. The film readout requires scanning and calculation of the OD. Also, to have consistent results, it is better to use the same film lot and scan the film after a fixed period post-irradiation. Nonetheless, film can provide valuable 2D profile of the beams at very high resolution.

For proton and heavy particle beams, the radiochromic film responses are sensitive at the high LET region due to its quenching effect[32, 33]. In this study, the calibration and small mice irradiations used the shallow depth plateau region of the conventional and high dose-rate proton beams. Caution must be taken when using the film at the beam Bragg peak region.

**CONCLUSION**

We have established an efficient proton dose calibration and monitoring procedure for small animal FLASH experiments. Using a Faraday cup is time-efficient and cost-effective than film dosimetry methods and can obtain the dose-rate information. The method is dose rate independent. We obtained a linear relationship between small-field average dose vs. Faraday cup reading for mice brain small field irradiation. For different collimation sizes, new calibration curves will need to be established. This method has been successfully used in ongoing pre-clinical mice proton experiments with both conventional and ultra-high dose rates. Similar approach can be adopted for small-field proton dosimetry for clinical beams.